\documentclass[aps,prl,twocolumn,showpacs,superscriptaddress,floatfix]{revtex4-1}
\usepackage{amsmath,amssymb}
\usepackage{graphicx}
\usepackage{epsfig}
\usepackage{color}
\usepackage{rotating}
\usepackage{bm}
\usepackage{setspace}
\begin{document}
\title{Direct Visualization of Ambipolar Mott Transition in Cuprate CuO$_2$ Planes}
\author{Yong Zhong}
\author{Jia-Qi Fan}
\author{Rui-Feng Wang}
\author{ShuZe Wang}
\affiliation{State Key Laboratory of Low-Dimensional Quantum Physics, Department of Physics, Tsinghua University, Beijing 100084, China}
\author{Xuefeng Zhang}
\affiliation{Institute of Physics, National Center for Electron Microscopy in Beijing, School of Materials Science and Engineering, Tsinghua University, Beijing 100084, China}
\author{Yuying Zhu}
\author{Ziyuan Dou}
\author{Xue-Qing Yu}
\author{Yang Wang}
\affiliation{State Key Laboratory of Low-Dimensional Quantum Physics, Department of Physics, Tsinghua University, Beijing 100084, China}
\author{Ding Zhang}
\affiliation{State Key Laboratory of Low-Dimensional Quantum Physics, Department of Physics, Tsinghua University, Beijing 100084, China}
\affiliation{Frontier Science Center for Quantum Information, Beijing 100084, China}
\author{Jing Zhu}
\affiliation{Institute of Physics, National Center for Electron Microscopy in Beijing, School of Materials Science and Engineering, Tsinghua University, Beijing 100084, China}
\author{Can-Li Song}
\email[]{clsong07@mail.tsinghua.edu.cn}
\author{Xu-Cun Ma}
\email[]{xucunma@mail.tsinghua.edu.cn}
\affiliation{State Key Laboratory of Low-Dimensional Quantum Physics, Department of Physics, Tsinghua University, Beijing 100084, China}
\affiliation{Frontier Science Center for Quantum Information, Beijing 100084, China}
\author{Qi-Kun Xue}
\email[]{qkxue@mail.tsinghua.edu.cn}
\affiliation{State Key Laboratory of Low-Dimensional Quantum Physics, Department of Physics, Tsinghua University, Beijing 100084, China}
\affiliation{Frontier Science Center for Quantum Information, Beijing 100084, China}
\affiliation{Beijing Academy of Quantum Information Sciences, Beijing 100193, China}

\begin{abstract}
Identifying the essence of doped Mott insulators is one of the major outstanding problems in condensed matter physics and the key to understanding the high-temperature superconductivity in cuprates. We report real space visualization of Mott insulator-metal transition in Sr$_{1-x}$La$_{x}$CuO$_{2+y}$ cuprate films that cover both the electron- and hole-doped regimes. Tunneling conductance measurements directly on the cooper-oxide (CuO$_2$) planes reveal a systematic shift in the Fermi level, while the fundamental Mott-Hubbard band structure remains unchanged. This is further demonstrated by exploring atomic-scale electronic response of CuO$_2$ to substitutional dopants and intrinsic defects in a sister compound Sr$_{0.92}$Nd$_{0.08}$CuO$_2$. The results may be better explained in the framework of self-modulation doping, similar to that in semiconductor heterostructures, and form a basis for developing any microscopic theories for cuprate superconductivity.
\end{abstract}

\maketitle
\begin {spacing}{0.99}
High-temperature superconductivity in cuprates develops upon doping a state of matter that is insulating due to strong electron-electron correlation of CuO$_2$ plane \cite{Lee2006doping,keimer2015quantum,kohno2018characteristics}. As Mott insulators, the ground state of cuprates could be characterized by a charge-transfer gap (CTG) between charge-transfer band (CTB) and upper-Hubbard band (UHB), derived predominantly from the O 2p and Cu 3d$_{x^2-y^2}$ orbitals, respectively \cite{ye2013visualizing,cai2016visualizing}. Whether the doping prompts spectral weight transfer from the high- to low-energy scale \cite{Meinders1993spectral} and pins Fermi energy ($E_\textrm{F}$) by some putative midgap states within CTG \cite{ye2013visualizing,cai2016visualizing,wu2013anomalous} or it induces $E_\textrm{F}$ shift of the Mott-Hubbard insulating state \cite{van1994electronic,Armitage2002doping,Shen2004missing} is a fundamental question but remains unresolved. The difficulty of settling this open issue is partially owing to specific doping level \cite{ye2013visualizing,Shen2004missing} or limited doping range of samples \cite{cai2016visualizing,Armitage2002doping} explored before, partially to the complex layered structure of cuprates. The latter makes direct experimental access to the CuO$_2$ planes that are sandwiched between the charge reservoir layers extremely challenging \cite{ye2013visualizing,cai2016visualizing,Armitage2002doping,Shen2004missing,fischer2007scanning}. Under this context, the systematic measurement of a cuprate system terminated by the CuO$_2$ planes, covering sufficiently broad doping range and both the electron ($n$)- and hole ($p$)-doped regimes \cite{segawa2010zero}, is the most effective way to solving the major problem.

We report such measurement by choosing an infinite-layer SrCuO$_2$, because, besides its simplest crystal structure among cuprates [Fig.\ 1(a), inset], its surface is terminated with CuO$_2$ \cite{Harter2012nodeless,Harter2015doping,Zhong2018atomic}. Given the metastability of bulk infinite-layer cuprates, an ozone-assisted molecular beam epitaxy technique is utilized to prepare single-crystalline films of Sr$_{1-x}$La$_{x}$CuO$_{2+y}$ (SLCO) and Sr$_{0.92}$Nd$_{0.08}$CuO$_2$ (SNCO) with well-controlled dopant type and concentration \cite{supplementary}. This enables to access a broad doping regime in the phase diagram and investigate the Mott physics on CuO$_2$ comprehensively.

Figure 1(a) shows a series of X-ray diffraction (XRD) patterns of SLCO films with various La doping level $x$. The good crystallinity is evident by the distinct Kiessig fringes in Fig.\ S1. Quantitative analysis of the XRD data reveals two distinct phases: a phase with $c$-axis lattice constant $c$ $\sim$ 3.45 $-$ 3.50 $\textrm{\AA}$ for $x <$ 0.104 (circles) and another with $c$ $\sim$ 3.65 $\textrm{\AA}$ for $x >$ 0.132 (squares), while for $x$ = 0.104 to 0.132 the two phases coexist [Fig.\ 1(b)]. Both phases belong to the family of infinite-layer cuprates sharing the same crystal structure, except for a small difference by  $\sim$ 0.2 $\textrm{\AA}$ in $c$ \cite{karimoto2001single, leca2006superconducting}. Yet, the Hall effect measurements reveal that the two phases exhibit $n$-type (electron) and $p$-type (hole) conductivity, respectively, involving a carrier-sign reversal at $x$ $\sim$ 0.104 $-$ 0.132 [Fig.\ S2]. Similar phenomenon was evidenced in electron-doped La$_{2-x}$Ce$_x$CuO$_4$ \cite{Sarkar2017fermi}, but the mechanism differs between them (Supplementary section 1). Thus we label the two phases as $n$-SLCO and $p$-SLCO, indicated by the blue and black arrows in Fig.\ 1(a), respectively.

The phase identification and atomically sharp interface in SLCO/SrTiO$_3$ heterostructures are further established by high-resolution scanning transmission electron microscopy (STEM). By applying the integrated differential phase contrast-STEM imaging technique \cite{lazic2016phase} that enables simultaneous visualization of all atoms, we show that the variation of $c$-axis lattice constant and concomitant carrier-sign reversal are due to the appreciable intake of apical oxygens as $x >$ 0.132 [Figs.\ 1(c) and 1(d)]. Thus, the oxygen stoichiometry in $p$-SLCO is significantly higher than that in $n$-SLCO. We also carry out \textit{in-situ} scanning tunneling microscopy (STM) imaging of the SLCO films. The $n$-SLCO films display a pristine CuO$_2$(1 $\times$ 1) surface with the anticipated Cu-Cu spacing of 3.9 $\textrm{\AA}$ [Fig.\ 1(e)], whereas the $p$-SLCO ones are characteristic of a (2 $\times$ 2) superstructure (white square) [Fig.\ 1(f)]. The (2 $\times$ 2) superstructure in $p$-SLCO is consistent with the periodic occupation of apical oxygens, whereas the integrity of CuO$_2$ is maintained (Supplementary section 2, Fig.\ 1(d) and Fig.\ S3). The excellent consistency among XRD, STEM, STM and Hall measurements indicates that a system for exploring the physics of doped Mott insulator directly on CuO$_2$ has been prepared.
\end {spacing}

\begin{figure}[t]
\includegraphics[width=\columnwidth]{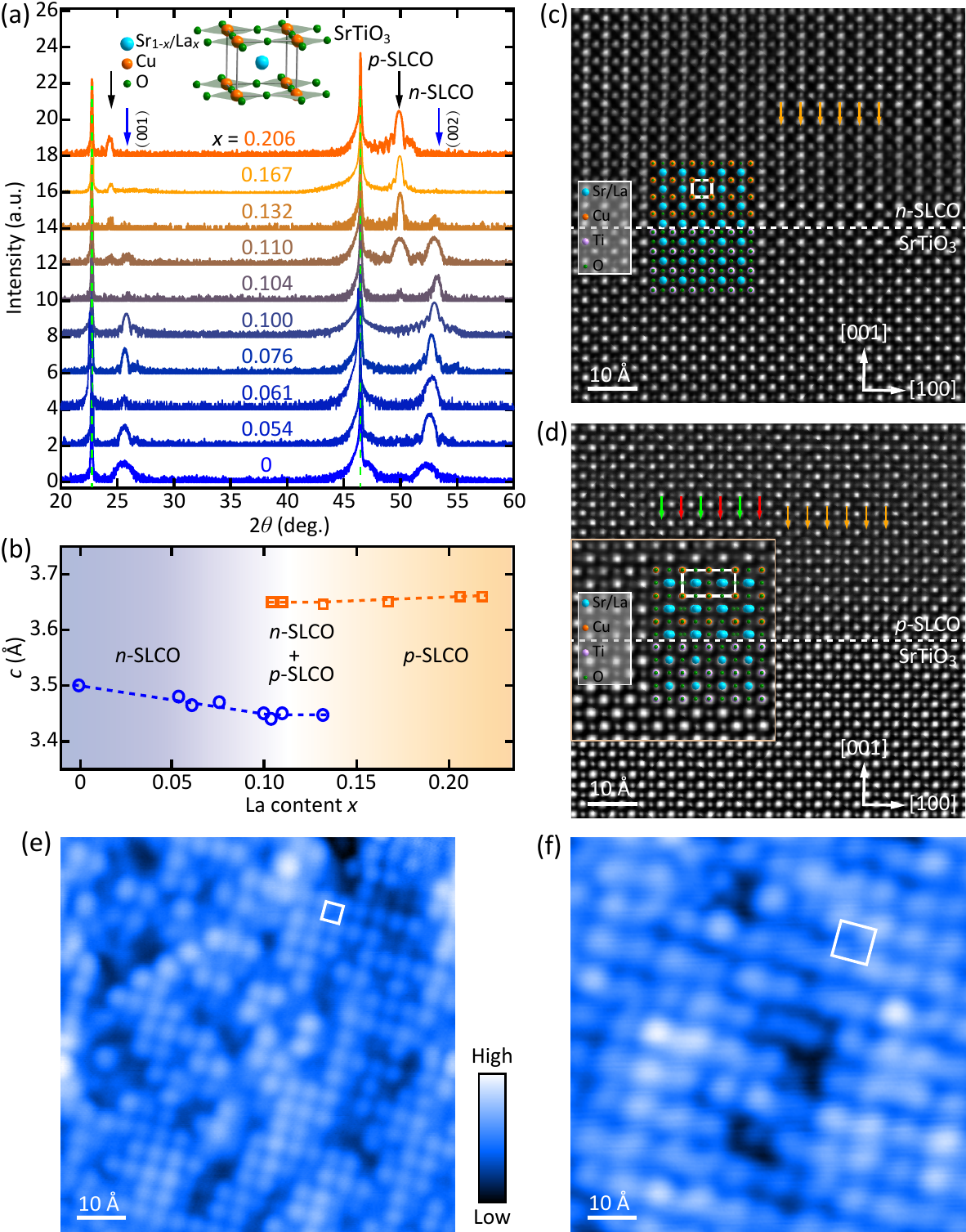}
\caption{(a) XRD spectra for Sr$_{1-x}$La$_{x}$CuO$_{2+y}$ films ($\sim$ 60 unit cells) at varied $x$. Inset shows the schematic crystal structure of SLCO. (b) Out-of-plane lattice constant $c$ as a function of $x$. (c, d) Integrated differential phase contrast-STEM images of $n$($p$)-SLCO/SrTiO$_3$ heterostructures, with the interfaces marked by the horizontal dashes. In contrast to $n$-SLCO without apical oxygen, an excess of oxygens (marked by the red and green arrows) register at the apical sites of Cu and change alternately in intensity and shape in $p$-SLCO cuprates, while the planar oxygens (orange arrows) in the CuO$_2$ planes exhibit no spatial variation. (e, f) STM topographies (80 $\textrm{\AA}$ $\times$ 80 $\textrm{\AA}$) of $n$-type ($x$ = 0.076) and $p$-type ($x$ = 0.167) SLCO films. Tunneling conditions are (e) $V$ = -1.0 V, $I$ = 10 pA and (f) $V$ = -0.3 V, $I$ = 50 pA.
}
\end{figure}

\begin{figure*}[h]
\includegraphics[width=1.48\columnwidth]{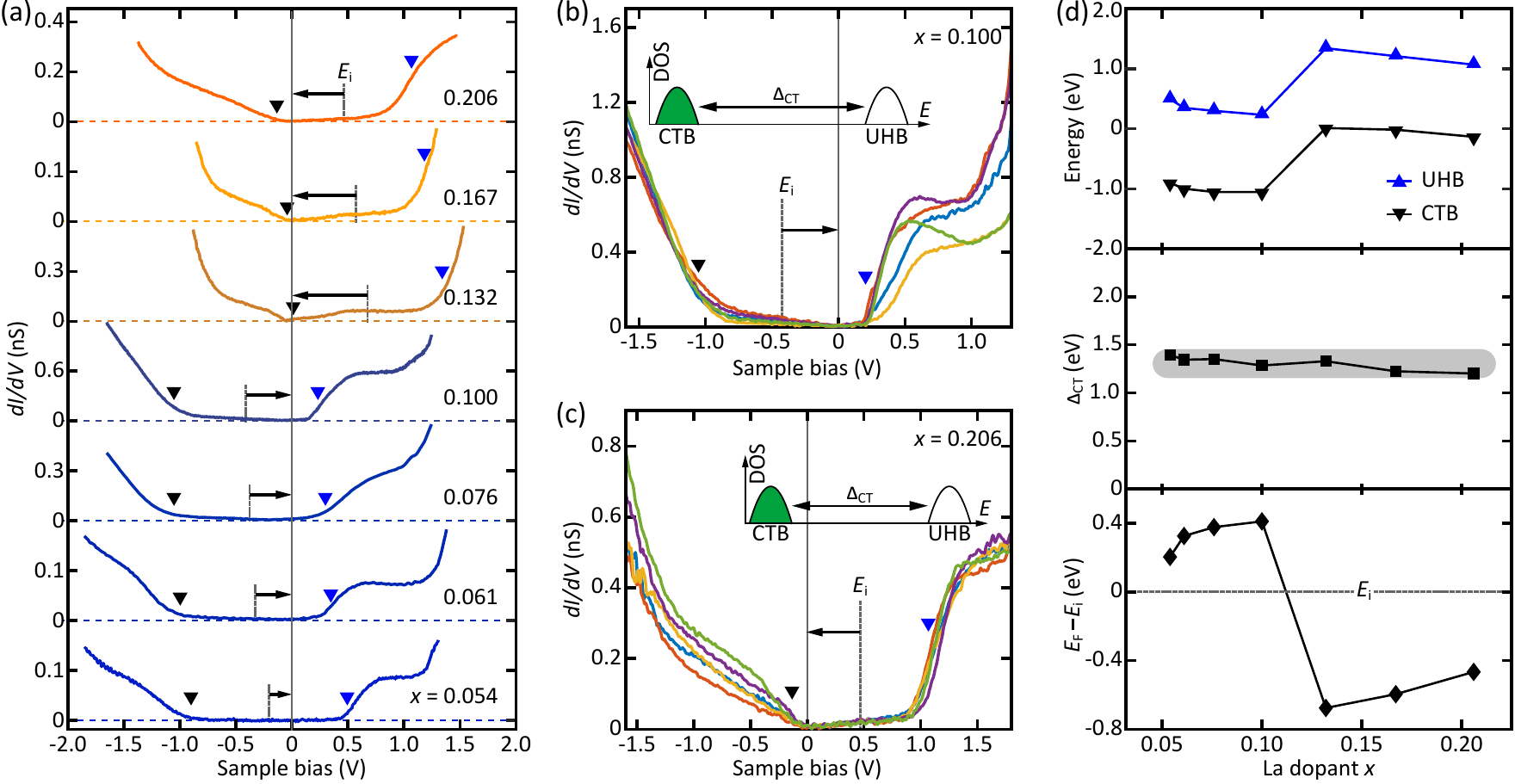}
\caption{(a) Spatially averaged differential conductance \textit{dI/dV} spectra of SLCO films under various doping. Gray solid lines and short bars indicate $E_\textrm{F}$ and the midgap energy $E_\textrm{i}$ throughout. For each \textit{dI/dV} spectrum, the tunneling junction is stabilized to the starting voltage from the negative side and $I$ = 100 pA except for $x$ = 0.132 and 0.167 ($I$ = 200 pA). (b, c) Representative \textit{dI/dV} spectra taken at equal separations (1 nm) in $n$-SLCO ($x$ = 0.100, $V$ = -1.6 V, $I$ = 100 pA) and $p$-SLCO ($x$ = 0.206, $V$ = 1.8 V, $I$ = 200 pA). Inserted are schematic energy bands of cuprates, only showing the CTB (green) and UHB (unfilled). (d) Statistically measured onset energies of CTB and UHB (top panel), charge-transfer gap $\Delta_{\textrm{CT}}$ (middle panel) and $E_\textrm{F}$ shift relative to $E_\textrm{i}$ (bottom panel) versus the La dopant.
}
\end{figure*}

\begin{figure*}[t]
\includegraphics[width=2\columnwidth]{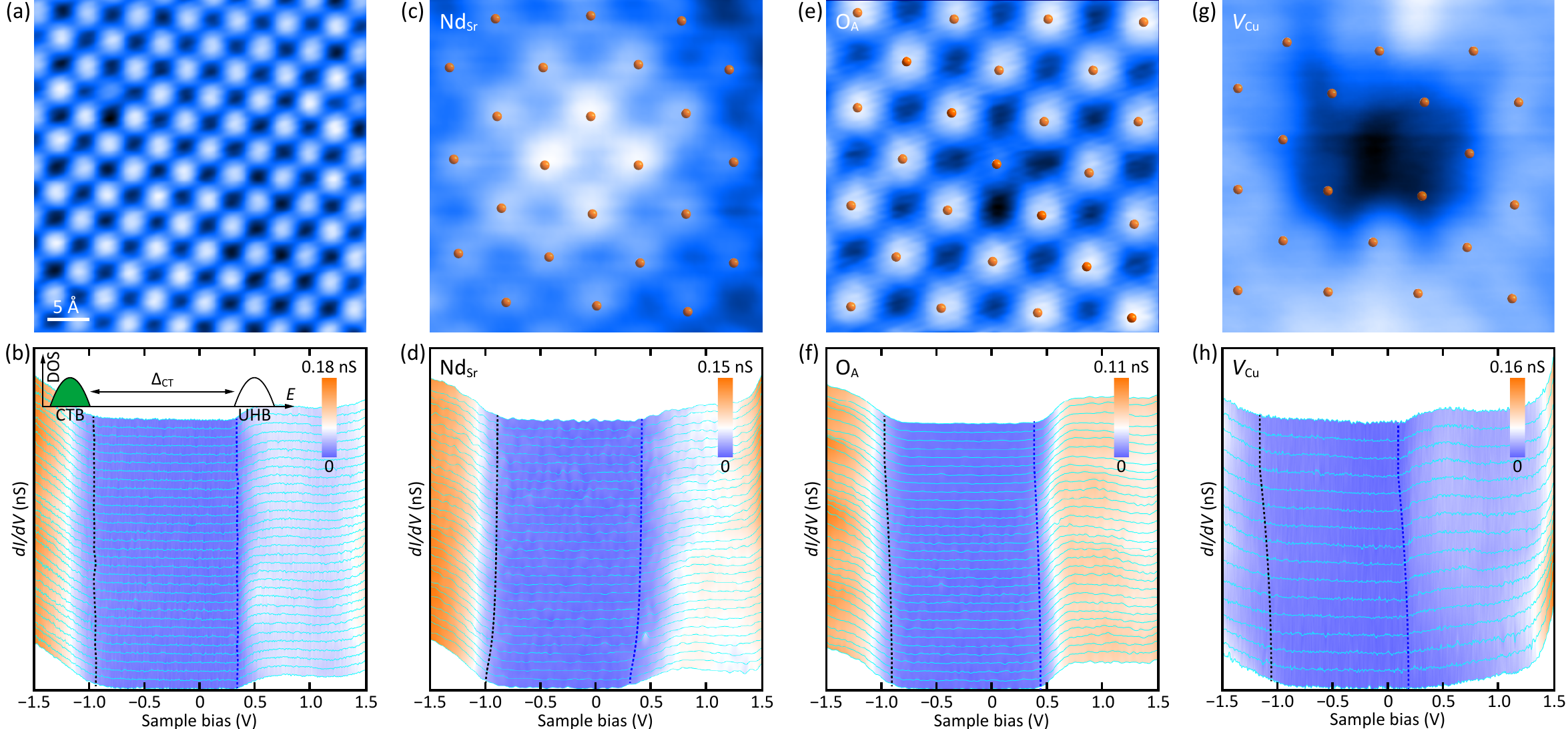}
\caption{(a) STM topography of SNCO films (40 $\textrm{\AA}$ $\times$ 40 $\textrm{\AA}$, $V$ = 1.5 V, $I$ = 20 pA). (b) Tunneling spectra acquired along a trajectory of 26.5 $\textrm{\AA}$ on CuO$_2$. The black and red dashes mark the UHB and CTB onsets, respectively. (c, d) Atomic-resolution topography (20 $\textrm{\AA}$ $\times$ 20 $\textrm{\AA}$, $V$ = -1.4 V, $I$ = 20 pA) of a single Nd$_{\textrm{Sr}}$ dopant and \textit{dI/dV} spectra along a trajectory of 16.4 $\textrm{\AA}$ from Nd$_{\textrm{Sr}}$ (bottom curve). (e, f) STM topography (20 $\textrm{\AA}$ $\times$ 20, $V$ = 1.0 V, $I$ = 20 pA) of an intrinsic O$_\textrm{A}$ acceptor and \textit{dI/dV} spectra along a trajectory of 12.2 $\textrm{\AA}$ from its center (bottom curve). (g, h) STM topography (20 $\textrm{\AA}$ $\times$ 20, $V$ = 1.5 V, $I$ = 20 pA) of a Cu vacancy (V$_{\textrm{Cu}}$) and \textit{dI/dV} spectra along a trajectory of 12.8 $\textrm{\AA}$ from its center (bottom curve). The setpoint is stabilized at $V$ = 1.5 V and $I$ = 50 pA. Orange dots denote the top Cu atoms.
}
\end{figure*}

Our most significant finding is that the fundamental Mott-Hubbard bands remain essentially unchanged, while $E_\textrm{F}$ systematically moves with doping. This is unambiguously revealed by measuring the electronic density of states (DOS) of CuO$_2$ planes at various doping levels via STM, as enumerated in Figs.\ 2(a)-(c) and Fig.\ S4. A comparison of the electronic DOS with the schematic band structure [Figs.\ 2(b) and 2(c)] shows the overall similar electronic structure:\ a CTG between CTB (its onsets are marked by black triangles) and UHB (its onsets are marked by blue triangles) is invariably present. Figure 2(d) depicts the onset energies of CTB and UHB as a function of the La doping $x$, from which the magnitude $\Delta_{\textrm{CT}}$ of CTG separating CTB and UHB is extracted. Here the CTB (UHB) onsets are determined from the intersections of linear fits to the data above and below the CTB (UHB) [Fig.\ S4(b)]. Except for the undoped ($x$ = 0) and slightly doped ($x$ = 0.020) samples [Fig.\ S5] for which the $\Delta_{\textrm{CT}}$ cannot be measured correctly by STM due to the tip-induced band bending \cite{battisti2017universality}, $\Delta_{\textrm{CT}}$ holds constant at 1.30 $\pm$ 0.07 eV, irrespective of either samples at various La doping [Fig.\ 2(d)] or a sample over various regions [Fig.\ S4(d)]. This value is close to the gap size of $\sim$ 1.5 eV measured by optical conductivity on infinite-layer cuprates \cite{Tokura1990CuO}. Furthermore, the bandwidths of CTB and UHB have a value of approximately 0.46 eV irrespective of doping [Fig.\ S6], which is in good agreement with the previous report of 0.41 eV \cite{Harter2015doping}. Altogether, we conclude that doping doesnot disrupt the fundamental band structure of the CuO$_2$ planes of SLCO we have studied. Such finding agrees with the stability of Mott-Hubbard bands with the Sr doping up to $x$ = 0.3 in La$_{2-x}$Sr$_x$CuO$_4$ \cite{Brookes2015stability}.

As anticipated, the midgap energy ($E_\textrm{i}$, vertical bars in Fig.\ S5) lies close to $E_\textrm{F}$ ($V$ = 0) in undoped SLCO. With increasing La doping from 0.020 to 0.100, $E_\textrm{F}$ moves gradually away from $E_\textrm{i}$ (see the right-pointing arrows in Figs.\ 2(a) and 2(b)) and approaches UHB, consistent with $n$-type doping. However, at $x$ = 0.132, $E_\textrm{F}$ suddenly jumps down below $E_\textrm{i}$ (see the left-pointing arrows in Figs.\ 2(a) and 2(c)) and passes CTB because of the intake of apical oxygens [Fig.\ 1(d)], signifying a transition to $p$-type doping. As the $x$ is further increased, $E_\textrm{F}$ shifts upwards [Figs.\ 2(a) and 2(d)], a consequence of increased La donors and reduced apical oxygens (Supplementary section 1). Given that the only difference between $n$-SLCO and $p$-SLCO cuprates is the type of dominant ionized dopants, and that all STM measurements are conducted on the CuO$_2$ planes, this systematic shift of $E_\textrm{F}$ with doping should be inherent to doped Mott insulators.

We argue that the above findings on CuO$_2$ are of fundamental importance and contrast sharply with both scenarios of $E_\textrm{F}$ pinning by the midgap states \cite{ye2013visualizing,cai2016visualizing,Meinders1993spectral,wu2013anomalous,battisti2017universality} and collapsing of the Mott-Hubbard ground state upon doping iridates \cite{delaTorre2015collapse}, which were distinctively measured on the charge reservoir layers. Instead, our results bear resemblance to the modulation doping of Al$_x$Ga$_{1-x}$As/GaAs semiconductor heterostructures \cite{dingle1978electron}, with the roles of the valence and conduction bands of undoped GaAs played by the CTB and UHB of chemically undisturbed CuO$_2$ in SLCO, respectively. A minor distinction is that the separation of ionized dopants (La and apical oxygens) in the Sr layers and free carriers in CuO$_2$ occurs in SLCO itself, for which it is best described as self-modulation doping \cite{Fu2018selfmodulation}. In this scheme, the intrinsic band structure of stoichiometric CuO$_2$ and GaAs does not alter with doping in a fundamental fashion \cite{Brookes2015stability,Ifflander2015local}, although the correlated states occur in CuO$_2$. The role of La (O) dopants is to provide mobile electrons (holes) that push $E_\textrm{F}$ of the CuO$_2$ planes upwards (downwards). They are consistent with our observations in Fig.\ 2.

The finding of unchanged Mott-Hubbard band structure of CuO$_2$ on doping is further confirmed on the atomic scale in Sr$_{0.92}$Nd$_{0.08}$CuO$_2$, which presents an essentially atomically flat CuO$_2$ [Fig.\ 3(a)] and spatially more uniform DOS [Figs.\ 3(b) and S7]. Evidently, the overall electronic structure, measured $\Delta_{\textrm{CT}}$ $\sim$ 1.28 $\pm$ 0.04 eV and bandwidths of CTB (0.43 $\pm$ 0.06 eV)/UHB (0.46 $\pm$ 0.14 eV) resemble SLCO prominently. The observations affirm the immunity of the Mott-Hubbard band structure of CuO$_2$ to the dopant type (La$^{3+}$, Nd$^{3+}$ and O$^{2-}$) in the intervening Sr planes (or equivalently the charge reservoir layers), echoing the self-modulation doping scenario.

The local response of electronic DOS to single dopants or impurities provides additional insight into the self-modulation doping scheme. By measuring the lateral registries with respect to the top Cu atoms, we identify substitutional Nd donors (Nd$_{\textrm{Sr}}$), intrinsic acceptors of apical oxygen (O$_\textrm{A}$), Cu vacancies (V$_{\textrm{Cu}}$) and  characterize their nearby electronic DOS in Figs.\ 3(c)-(h). Around the Nd$_{\textrm{Sr}}$ donors, both CTB and UHB are shifted downwards [Fig.\ 3(d)], whereas the O$_\textrm{A}$ and V$_{\textrm{Cu}}$ acceptors locally move the bands upwards [Figs.\ 3(f) and 3(h)]. This band bending agrees with screened Coulomb potential (Fig.\ S8 and Supplementary section 3). Although a V$_{\textrm{Cu}}$ acceptor suppresses significantly UHB on a length scale of 4.3 $\textrm{\AA}$ [Figs.\ 3(h) and S8(d)], the whole Mott-Hubbard bands of CuO$_2$ are robust against Nd$_{\textrm{Sr}}$ and O$_\textrm{A}$ in the Sr charge reservoir layers [Figs.\ 3(d) and 3(f)].

\begin{figure}[t]
\includegraphics[width=\columnwidth]{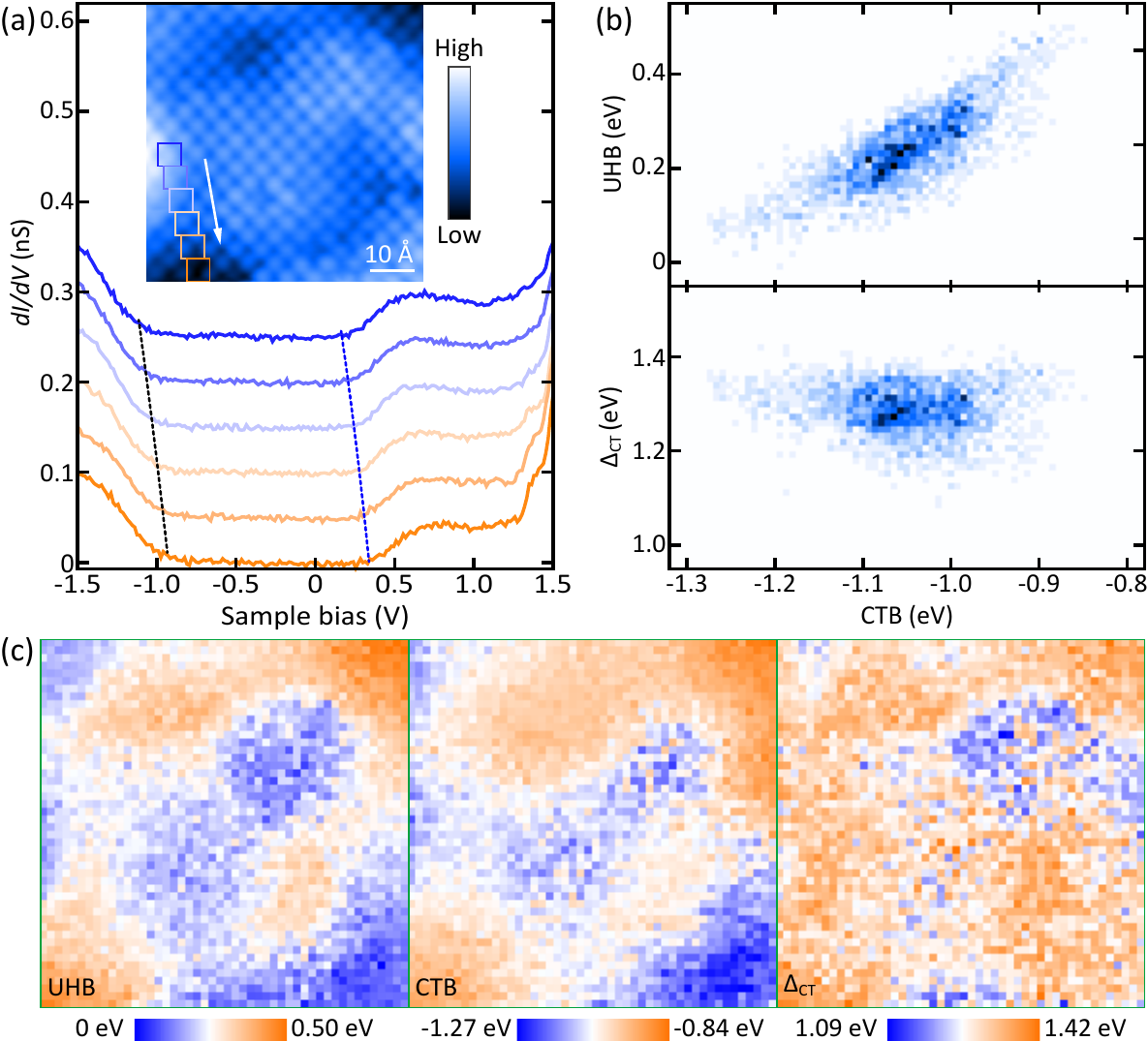}
\caption{(a) Local DOS spectra along the white arrow of the inserted STM topography (62 $\textrm{\AA}$ $\times$ 62 $\textrm{\AA}$, $V$ = 0.4 V, $I$ = 20 pA). Every curve is the mean of 16 \textit{dI/dV} spectra inside the color-coded square. For clarity the curves have been vertically offset by 0.05 nS. (b) Correlations between UHB onsets (top panel), $\Delta_{\textrm{CT}}$ (bottom panel) and CTB onsets. (c) Maps of UHB onsets, LHB onsets and $\Delta_{\textrm{CT}}$.
}
\end{figure}

The unchanged Mott-Hubbard band structure of infinite-layer cuprates against the defects of charge reservoir layers is corroborated by acquiring the 48 $\times$ 48 grid \textit{dI/dV} spectroscopy data over a field of view of 62 $\textrm{\AA}$ $\times$ 62 $\textrm{\AA}$, devoid of any defects in the topmost CuO$_2$ plane [Fig.\ 4(a), inset]. The Nd$_{\textrm{Sr}}$ donors and O$_\textrm{A}$ acceptors underneath are spatially distributed randomly that induce the contrast in STM image. As revealed, the electronic DOS displays spatial variation in the onsets of UHB and CTB, but the overall Mott-Hubbard band structure changes little [Fig.\ 4(a)]. This is more quantitatively proved in Figs.\ 4(b) and 4(c). On the bright regions, the UHB and CTB onset energies, which correlate positively with each other (top panel of Fig.\ 4(b)), are relatively lower, matching with the aggregation of Nd$_{\textrm{Sr}}$ donors there. The tiny spatial variation in $\Delta_{\textrm{CT}}$ [Fig.\ 4(c)] might arise from either the measurement uncertainty or band tailing associated with heavy doping \cite{Van1992theory}, or both.

Followed by the systematic shift in $E_\textrm{F}$, we observe dopant-induced in-gap states (IGS) that overspread the whole CTG [Figs.\ 2(a)-2(c), Fig.\ 3 and Fig.\ 4(a)] and induce a Mott insulator-metal transition as $x\geq$ 0.054 (Supplementary section 4 and Fig.\ S9). Figure S10 plots $x$-dependent lower-energy-scale conductance spectra near $E_\textrm{F}$. Intriguingly, smoothly varying electronic DOS with nanoscale puddles of pseudogap [Fig.\ S10(b)] are spatially separated by regions that are relatively featureless [Fig.\ S10(c)]. The pseudogaps exhibit pronounced electron-hole asymmetry that awaits further explanation (Supplementary section 4 and Fig.\ S11). In contrast to earlier STM studies on charge reservoir layers (e.g.\ BiO, CaCl and SrO) in cuprate compounds \cite{ye2013visualizing,cai2016visualizing,kohsaka2012visualization} and iridates \cite{battisti2017universality,okada2013imaging}, no peak- or hump-like electronic DOS has been observed within the CTG of CuO$_2$ planes. This finding is not expected from the emergent IGS associated with spectral weight transfer from the high- to the low-energy scale \cite{Meinders1993spectral}. From a view point of modulation-doping scheme in semiconductor physics \cite{dingle1978electron}, the dopant-induced continuum of IGS could be better explained invoking the confined two-dimensional electron/hole gas at the Sr$_{1-x}$(La, Nd)$_x$/CuO$_{2+y}$ interfaces, evanescent states within the charge-transfer gap of CuO$_2$ \cite{Ifflander2015local}, or a combination of them. The present experimental data are not sufficient to disentangle both scenarios, and the nature of emergent IGS responsible for the low-lying physics remains an open question in cuprates.

To conclude, our measurement of Mott insulator-metal transition in infinite-layer cuprates has presented several indispensable and exceptionally new results. These results are obtained from the direct measurement of the key CuO$_2$ plane, which is different from earlier studies that were usually conducted on the charge reservoir layers of cuprates.
The observed robust fundamental Mott-Hubbard bands against doping and the self-modulation doping-driven systematic shift of $E_\textrm{F}$ should form a starting point for developing any microscopic models of the Mott physics as well as the superconductivity mechanism in cuprates. Given that the superconductivity exclusively occurs in the CuO$_2$ planes for all cuprates, we believe that such a model is also applicable to other cuprates, which merits a future study.

\begin{acknowledgments}
We thank Y.\ Y.\ Wang, F.\ C.\ Zhang, G.\ M.\ Zhang, D.\ H.\ Lee and X.\ C.\ Xie for fruitful discussion. The work was financially supported by the Ministry of Science and Technology of China, the National Natural Science Foundation of China, and in part by the Beijing Advanced Innovation Center for Future Chip (ICFC). Y.\ Zhong, J.\ Q.\ Fan, and R.\ F.\ Wang contributed equally to this work.
\end{acknowledgments}

%\bibliography{ILcuprate}
%

\end{document}